\documentclass[a4paper]{article}

\usepackage{INTERSPEECH2021}
\newcommand{\etal}{\textit{et al.}}
\usepackage{hyperref}
\usepackage{xcolor}

\title{Emotional Prosody Control for Speech Generation}
\name{$^{*}$Sarath Sivaprasad$^{1,2}$, $^{*}$Saiteja Kosgi$^1$, Vineet Gandhi$^1$ \thanks{* equal contribution}}
\address{
  $^1$CVIT, KCIS, IIIT Hyderabad\\
  $^2$TCS Research, Pune}
\email{sarath.s@research.iiit.ac.in, saiteja.k@research.iiit.ac.in}

\begin{document}

\maketitle
\begin{abstract}
Machine-generated speech is characterized by its limited or unnatural emotional variation. Current text to speech systems generates speech with either a flat emotion, emotion selected from a predefined set, average variation learned from prosody sequences in training data or transferred from a source style. We propose a text to speech(TTS) system, where a user can choose the emotion of generated speech from a continuous and meaningful emotion space (Arousal-Valence space). The proposed TTS system can generate speech from the text in any speaker's style, with fine control of emotion. We show that the system works on emotion unseen during training and can scale to previously unseen speakers given his/her speech sample. Our work expands the horizon of the state-of-the-art FastSpeech2 backbone to a multi-speaker setting and gives it much-coveted continuous (and interpretable) affective control, without any observable degradation in the quality of the synthesized speech. Audio samples are available at {\scriptsize{\url{https://researchweb.iiit.ac.in/~sarath.s/emotts/}}}

\end{abstract}
\noindent\textbf{Index Terms}: speech generation, prosody control, human-computer interaction

\section{Introduction}

Text-to-speech(TTS) applications strive to synthesize `human-like speech.' This task not only needs modeling of the human vocal system (to generate the frequencies given a sequence of phonemes), but also captures the prosody and intonation variations present in human speech. Neural network models have made significant improvements in enhancing the quality of generated speech, and most state-of-the-art TTS systems like Deep Voice\cite{arik2017deep}, Tacotron\cite{wang2017tacotron}, and Fastspeech2\cite{ren2020fastspeech} generate natural sounding voice. However, high-level affective controllability remains a much-coveted property in these speech generation systems and is a problem of interest in the speech community for well over three decades~\cite{cahn1990generation,schroder2001emotional}.

Controlling emotional prosody (affective control) is vital for many creative applications (like audiobook generation, virtual assistants) and desirable in almost all speech generation use cases. Affective control is a challenging task, and even with the significant improvements in recent years, TTS systems today do not have high-level interpretable emotion control. The existing systems are restricted to either transfer of prosody from source style\cite{karlapati2020copycat} or learning prosody globally given a phoneme sequence\cite{ren2020fastspeech}. Habib~\etal~\cite{habib2019semi} proposed a system to control affect; however, their method cannot incorporate fine control and is limited to six discrete emotional states. 

We propose a TTS system based on FastSpeech2\cite{ren2020fastspeech} and bring in fine-grain prosody control and multi-speaker control. The improvements are achieved without any observable degradation in the synthesized speech quality and without compromising its ultra-fast inference. Like Fastspeech2, our model predicts low-level features from the phoneme sequence (e.g., pitch, energy, and duration). However, the proposed model incorporates high-level and interpretable sentence-level control over the low-level intermediate predictions computed for each phoneme. Our approach has profound implications from a usability perspective because (a) for a human, a phoneme level control is difficult to interact with, and our model allows sentence level emotional control and (b) low-level features like pitch, energy, duration, etc. are difficult to interpret and by conditioning them on \emph{arousal valence} values, our model allows an expressible emotional control. We condition the encoder to scale for multiple speakers and transform the encoded vector to incorporate the continuous \emph{arousal-valence} values. Our core contributions are:

\begin{itemize}
\item We extend the FastSpeech2 architecture to scale for multiple speakers based on fixed-size speaker embeddings.
\item We propose a novel Prosody Control(PC) block into FastSpeech2 architecture to incorporate high-level affective sentence level control. We use scalar \emph{Arousal-Valence} values on the low level and phoneme level variance features like pitch, energy, and duration. 
\item The proposed architecture hence allows to generate speech with fine grain emotional control as they can be chosen from a continuous and interpretable \emph{Arousal-Valence} space. 
\end{itemize}

\begin{figure*}[t]
  \centering
  \includegraphics[width=\linewidth]{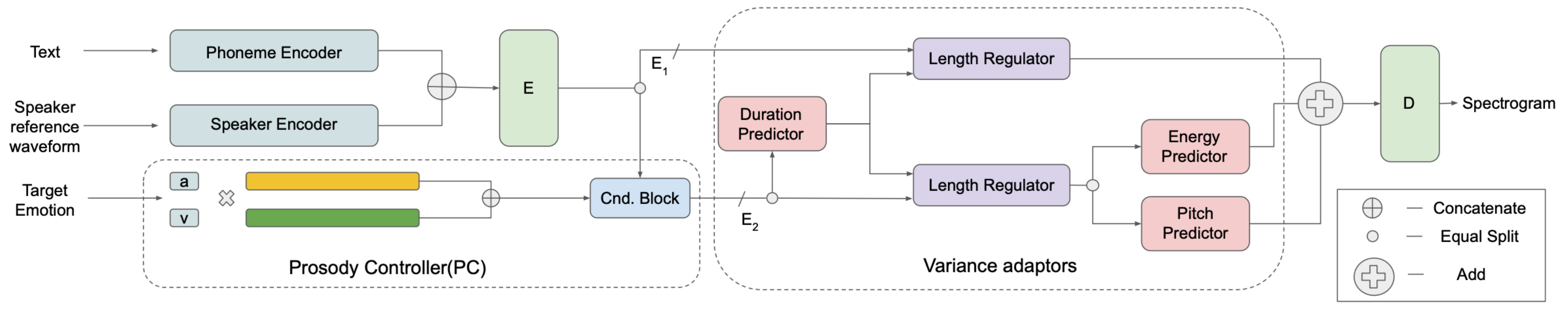}
  \caption{Schematic diagram of the proposed model.}
  \label{fig:speech_production}
\end{figure*}

\section{Related work}

{\bf Neural TTS:} Neural network-based TTS have changed the landscape of speech synthesis research and have significantly improved the speech quality over conventional concatenative and statistical parametric approaches~\cite{hunt1996unit,wu2016merlin}. Some of the recent popular neural TTS systems are Tacotron~\cite{wang2017tacotron}, Tacotron 2~\cite{shen2018natural}, Deep Voice 1,2,3~\cite{arik2017deep,arik2017deep2} and ClariNet~\cite{ping2018clarinet}. These approaches first generate Mel-spectrogram autoregressively from text input. The Mel-spectrogram is then synthesized into speech using vocoders like Griffin-Lim~\cite{griffin1984signal}, WaveNet~\cite{oord2016wavenet} and Parallel WaveNet~\cite{oord2018parallel}. More recently, the FastSpeech~\cite{ren2019fastspeech} and FastSpeech 2~\cite{ren2020fastspeech} methods approach TTS in a non-autoregressive manner and show extremely high computational gains during training and inference. Despite synthesizing natural-sounding speech, the above-mentioned neural TTS models give little or no control over the emotional expression for a given sentence. \\

\noindent {\bf Multiple Speaker TTS:} There has been a major focus on scaling TTS systems to multiple speakers. Early neural multi-speaker TTS models require tens of minutes of training data per speaker. Fan~\etal~\cite{fan2015} proposed a neural network model which uses a shared hidden state representation for multiple speakers and speaker-dependent output layers. Gibiansky~\etal~\cite{arik2017deep2} introduced a multi-speaker variation of Tacotron, which learned low-dimensional speaker embeddings for each training speaker. Their later work~\cite{ping2018deep} scaled up to support over 2,400 speakers. Such systems~\cite{arik2017deep2,ping2018deep,fan2015} learn a fixed set of speaker embeddings and therefore only support the synthesis of voices seen during training. More recent approaches decouple speaker modeling from speech synthesis by independently training a speaker-discriminative embedding network~\cite{nagrani2017voxceleb}. The TTS models are then conditioned on these speaker-discriminative embedding obtainable from a few seconds of speech for the given speaker. Wan~\etal~\cite{wan2018generalized} train speaker verification network, Jia~\etal~\cite{jia2018transfer} condition the Tacotron 2 model on the embeddings of verification network.  Our work extends such zero-shot multi-speaker support for a non-autoregressive model, FastSpeech2. \\

\noindent{\bf Prosody Control:} Following enormous progress in neural TTS systems, the focus in recent years has shifted to modeling latent aspects of prosody. Humans speak with different styles and tonal variations, but there is an underlying pattern or constraint to these varying styles. The absence of an expected variation or the presence of an unexpected variation is easily detected as an uncanny speech by a human listener.

Wang~\etal~\cite{wang2018style} proposed a framework to learn a bank of style embeddings called ``Global Style Tokens" (GST) that are jointly trained within Tacotron (without any explicit supervision). A weighted combination of these vectors corresponds to a range of acoustic variations. Battenberg~\etal~\cite{battenberg2019effective} introduce a hierarchical latent variable model to separate style from prosody. Although such unsupervised methods~\cite{wang2018style,battenberg2019effective} can achieve prosodic variations, they can be hard to interpret and do not allow a straightforward control for varying the emotional prosody.

Skerry-Ryan~\etal~\cite{skerry2018towards} proposed an end-to-end framework for prosody transfer, where the representation of prosody is learned from reference acoustic signals. The system transfers prosody from one speaker to another in a pitch-absolute manner. Karlapati~\etal~\cite{karlapati2020copycat} proposed a framework for reference prosody by capturing aspects like rhythm, emphasis, melody, etc., from the source speaker. However, such reference-based methods cannot give a desired level of control as it requires a source reference for each different style of utterance. While such methods can work for scenarios like dubbing, they fall short on audiobook generation and other creative applications.

\begin{figure}[t]
  \centering
  \includegraphics[width=0.85\linewidth]{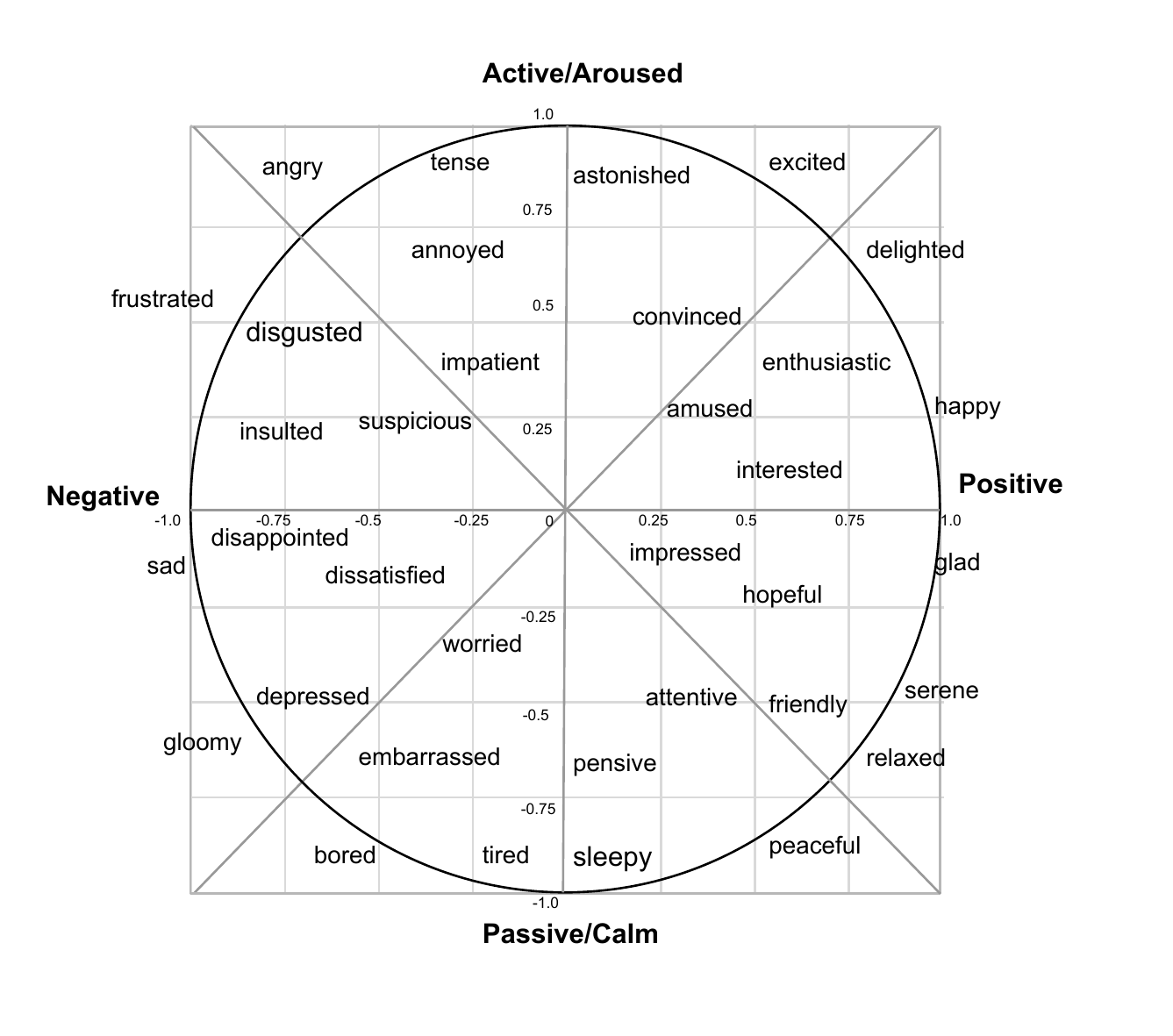}
  \caption{The 2-D Emotion Wheel.}
  \label{fig:avmodel}
\end{figure}

Habib~\etal~\cite{habib2019semi} proposed a generative TTS model with a semi-supervised latent variable that can control affect in discrete levels. Data collection involved recording reading text in either a happy, sad or angry voice at two levels of arousal. These six levels of arousal-valence combinations were used for partial supervision of latent variables. The model brings control only over discrete affective states (6 points), only representing a subset of emotions. Our work extends this idea by giving affect control over the continuous space of arousal and valence. Arousal(A) is a measure of intensity, whereas Valance (V) describes emotion's positivity or negativity. Russell~\etal~\cite{russell1980circumplex} show that these two parameters can represent various emotions in a 2D plane (Figure ~\ref{fig:avmodel}).

By conditioning TTS on AV values, our work allows fine-grained and interpretable control over the synthesized speech. We choose Fastspeech2\cite{ren2020fastspeech} as the backbone due to its simplicity and ultra fast inference speed. Fastspeech2 predicts low level features like pitch, duration and energy for each phoneme and conditions the decoder on them. Our work facilitates a sentence level conditioning of these phoneme level features using scalar values for Arousal-Valence (AV).

\section{Model}
Our model uses Fastspeech2 as its backbone\cite{ren2020fastspeech}. Unlike autoregressive models, Fastspeech2 does not depend on the previous frames to generate next frames, leading to faster synthesis. The model comprises of mainly three parts, namely: the encoder-decoder block, the prosody control block, and variance adaptor (Figure~\ref{fig:speech_production}). The encoder block(E) and decoder block(D) are feed-Forward transformers with self-attention and 1-D convolution layers. 
The model has three inputs: \begin{itemize}
    \item Text: The text to be rendered as speech
    \item Speaker reference waveform: Audio sample of source speaker in whose voice the output will be rendered.
    \item Target Emotion: Arousal and valence values corresponding to the target emotion
\end{itemize}

Phoneme encoder gives a vector representation of fixed size for each of the phonemes in the input text. These embeddings are padded along sequence dimensions to match the number of phonemes in all the inputs across a batch. To incorporate speaker information into these embeddings, we condition our encoder(E) on speaker identity by using an embedding trained for speaker verification \cite{wan2018generalized}. These embeddings capture the characteristics of different speakers, invariant to the content and background noise. Given a speaker reference waveform, using the pre-trained model, we generate 256 dimension speaker embedding. We concatenate the phoneme embedding with speaker embedding along the sequence dimension at the zeroth position. i.e., the speaker embedding appears as the first phoneme in the concatenated vector. This technique ensures the constant position of speaker embedding (irrespective of pad length of phonemes). The encoder(E) learns a representation for each phoneme attending to all other phonemes along with speaker embedding. We call this representation as $E_1$. We observed that conditioning the encoder with speaker embedding gives better results than conditioning the decoder with speaker embedding. In our model, conditioning the decoder with speaker embedding did not capture the speaker's identity. We hypothesize this is because the variance predictions are dependent on speaker embeddings. The encoder's output and predicted variances (pitch, energy, duration) are decoded(at D) to obtain the Mel-spectrogram. The loss is computed between the generated Mel-spectrogram and the spectrogram of target speech(Mel-loss). This end-to-end structure forms the backbone of our system.  

The Prosody Control(PC) block generates latent representation for each phoneme with affective cues from arousal and valence. We use two learnable vectors of length 256 to represent arousal and valance, respectively. The combined emotion is computed as the sum of these two vectors, weighted by arousal and valence inputs. The two vectors are trained with the loss computed at each of the variance predictors along with Mel-loss. The weighted sum is concatenated with $E_1$ and passed through a linear layer(condition block). The resulting representation is a phoneme embedding incorporating input emotions. We call this representation as $E_2$. 

$E_2$ is passed through the duration predictor, which predicts a duration for each of the phonemes. Based on the duration ($d$) predicted for each phoneme using $E_2$, the length regulator expands the hidden states of the phoneme sequence $d$ times for both $E_1$ and $E_2$. The total length of the hidden states in the two regulated embeddings now corresponds to the length of the output Mel-spectrograms.  Pitch and energy are predicted at corresponding variance predictors using regulated $E_2$. Each variance predictor is trained with corresponding ground truth extracted from the speech wave. The energy and pitch computed are added to regulated $E_1$ and are passed to decoder block(D). The decoder outputs the Mel spectrogram. We use the MelGan vocoder~\cite{kumar2019melgan} to generate raw speech from the spectrogram. 

 We use $E_2$ to predict the variances and use the resultant predictions to modify $E_1$. Decoder gets $E_1$ as input, which is not concatenated with affective cues. This strategy ensures that the emotion only modifies the pitch, energy, and duration, and the encoder-decoder module of the TTS can be trained, independent of the prosody control block. We propose this strategy to train the backbone and prosody controller block on LibriSpeech and MSP datasets independently. This ensures that errors incurred in transcribing MSP does not effect TTS quality. We train the prosody control block separately after training and freezing the encoder-decoder modules. 

\section{Experiments and Results}

\subsection{Dataset}
We use two datasets to train our model. We train our backbone multi-speaker TTS model (leaving out Prosody Controller block) on LibriSpeech~\cite{libritts} dataset. LibriSpeech~\cite{libritts} contains transcripts and corresponding audio samples spoken by multiple speakers. Our model takes phoneme sequences as input. The text sequence is converted into phoneme sequence using the method proposed in~\cite{sun2019token}. We generate Mel spectrogram from the audio file following the work in~\cite{shen2018natural}. This is used to compute loss with predicted Mel spectrogram. We compute energy, pitch, and duration from the speech to train the corresponding variance predictors. To train the duration predictor, we generate ground truth values of duration per phoneme using Montreal Forced aligner(MFA)~\cite{mfa}. MFA is a speech-text alignment tool used to generate time-aligned versions of audio files from the given transcript. LibriSpeech consists of two clean training sets comprising 436 hours of speech training data. We train on this data and use some of the speaker samples as a validation set. This dataset has no affective annotations.

We train our Prosody Controller(PC) block on MSP Podcast corpus~\cite{Lotfian_2019_3} . MSP Podcast is a speech corpus annotated with emotions. It consisting of podcast segments annotated with emotion labels and valance arousal values ranging from 1 to 7.  The corpus consists of 73K segments comprising ~100 hours of speech, split into training and validation data. MSP Podcast corpus does not contain transcripts for the audio segments. To generate transcripts, we use Google speech-to-text API. We use Montreal-Forced-aligner(MFA)~\cite{mfa} to achieve alignment, and if MFA does not find proper alignment for the text and audio pair, the sample is discarded. This accounts for the inaccuracies of the speech-to-text API and background noise in audio samples. After applying MFA and discarding the wrongly transcribed samples, we are left with 55k samples comprising roughly ~71 hours of speech. We use this data to train our prosody control module. 

\subsection{Training}
We train our model in two stages. 
\begin{itemize} 
    \item We first train our multi-speaker model barring Prosody Controller block on LibriSpeech~\cite{libritts} dataset. The encoder-decoder model with variance adaptors is trained together. The total loss consists of Mel loss(computed between predicted spectrogram and the spectrogram of corresponding ground truth audio), pitch loss, energy loss, and duration loss (each of which is computed directly from the ground truth audio). In this phase, $E_1$ is directly used as input to variance predictors. The model is trained on 4 GPUs with a batch size of 16. We use Adam optimizer to train the model. The training takes around 200k steps until convergence.
    \item In the second phase, we train our Prosody Controller block using MSP Podcast\cite{Lotfian_2019_3} corpus. We bring the emotion control by conditioning variance predictors on arousal valance values along with phoneme sequences($E_2$). In this phase, we freeze the weights of the encoder-decoder model trained on LibriSpeech and only train the PC and variance adaptors. The model is trained on 4 GPUs with a batch size of 16, and it takes 150k steps until convergence. 
\end{itemize} 

\begin{table}[t]
  \caption{The MOS with 95\% confidence intervals. }
  \label{tab:word_styles}
  \centering
  \begin{tabular}{ll}
    \toprule
    \textbf{Model}      & \textbf{Mean Opinion Score(MOS)}                \\
    Fastspeech2                    & $3.65 \pm 0.09$                                      \\
    
    Our Model              & $3.62 \pm 0.13$                               \\
    \midrule
    
     \textbf{Speaker Similarity}              &  \textbf{Mean Opinion Score(MOS)}                          \\
    Same speaker set                    & $3.6   \pm 0.08$                            \\
    
    Same gender speaker set          & $2.55   \pm 0.09$                 \\
    Different gender speaker set                & $1.2  \pm 0.04$           \\
    \midrule
    
     \textbf{Affect control}           &  \textbf{Avg. rater score in \%}           \\
    Superlative emotion match                    & 86.0                                                                  \\
    \bottomrule
  \end{tabular}
\end{table}
\subsection{Model Performance}

We measure the naturalness of generated speech, speaker sensitivity, and emotion control of our model through three user studies. We assess the voice's naturalness, speaker similarity using the Mean Opinion Score (MOS) collected from subjective listening tests. We use a Likert scale, with a range of 1 to 5 in 1.0 point increments. We evaluate emotion control using the average rater score. The results are reported in Table \ref{tab:word_styles}. The qualitative audio samples are available at {\scriptsize{\url{https://researchweb.iiit.ac.in/~sarath.s/emotts/}}}

\noindent {\bf Naturalness of generated speech:}
To evaluate the naturalness of the generated speech, we use a set of 30 phrases that do not appear in training set of either MSP or LibriSpeech and synthesize audio using our model. To compare the MOS of our model, we also synthesize the same phrases using Fastspeech2~\cite{ren2020fastspeech}. A collection of samples from both these models are provided to users. Twenty proficient English speakers are asked to make quality judgments about the naturalness of the synthesized speech samples and asked to rate on a Likert scale of range 1 to 5 where 1 being `completely unnatural' and 5 being `completely natural'. The results in Table \ref{tab:word_styles} show that similar scores are obtained for the two models. The results demonstrate that our model does not bring any noticeable distortions in terms of the naturalness of generated speech compared to the Fastspeech2 backbone. \\  

\noindent {\bf Capturing reference speaker voice :} Speaker similarity is evaluated in a similar fashion using MOS. We validate the speaker similarity on three different sets. 
\begin{itemize}
    \item Same speaker set:  This set consists of sample pairs synthesized from the same speaker. The pair consists of either a ground truth speech and a synthesized sample or both synthesized samples.
    \item Same-gender speaker set: Here, we synthesize phrases for a set of speakers of the same gender. We form pairs of samples with the same gender but different speaker.
    \item Different gender speaker set: This set is curated by pairing synthesized audio samples generated for speakers from opposite genders.
\end{itemize}

Given a pair of samples, participants were asked to rate the similarity score of how close the voices sound on a Likert scale of 1 to 5. Where 1 corresponds to `Not at all similar' and 5 corresponds to `extremely similar'. For the same speaker set, we obtained a MOS of  3.6. This shows that our model can synthesize voices that sound close to a given target speaker. The MOS of 2.55 for the same gender set shows that audio generated from different speakers of the same gender has a certain degree of similarity. Furthermore, the low MOS of 1.2 for samples from different genders shows that our model's synthesized speech can be discriminated based on gender.

\noindent {\bf Affective control:} Interpreting affect in rendition is subjective, challenging, and highly correlated with the content. We use user ratings to evaluate affect control. The model being conditioned on the continuous and meaningful space of emotion, user can change the level of emotion like happy to delighted, sad to depressed, etc., superlatively, during synthesis. We synthesize a set of phrases with different arousal valence(AV) values to evaluate the control obtained by changing AV values. 

For our survey, we choose samples consisting of different levels of four emotions: happy, sad, angry, and excited. We provide a pair of samples for each of the above emotions, with one sample corresponding to the lower level of the emotion and the other corresponding to the higher level of respective emotion (e.g., happy to delighted). We choose appropriate AV values such that particular emotion is expressed in two degrees. Raters are asked not to judge the content and choose the sample expressing the particular emotion strongly (e.g., which one is angrier or happier). Every rater is shown eight different pairs of samples. We choose two pairs from each of the aforementioned emotions. The reported score shows the average percentage score obtained by raters in choosing the stronger emotion.

Compiling results from all the users, we observe that 86\% of the raters can correctly choose the sample strongly expressing a particular emotion. The study shows the ability of the model to control prosody using arousal valance values. 

\section{Conclusions}
Our work addresses the problem of emotional prosody control in machine-generated speech. In contrast to previous prosody control methods, which are either difficult to interpret by humans, require reference audio or allow selection only among a discrete set of emotions, our method allows a continuous and interpretable variation. We use the FastSpeech2 TTS model as a backbone and add a novel Prosody Control (PC) block. The PC blocks conditions the phoneme level variational parameters on sentence-level Arousal Valance values. We also extend the FastSpeech2 framework to support multiple speakers by conditioning it on a discriminative speaker embedding. Our user study results demonstrate the efficacy of the proposed framework and show that it can synthesize natural-sounding speech, mimic reference speakers, and allow interpretable emotional prosody control. 

\section{Acknowledgement}
We want to thank Anil Nelakanti for the initial comments and discussions. The authors would also like to thank Carlos Busso and others at The University of Texas at Dallas for sharing the MSP-Podcast database. The work would not have been possible without this valuable resource. We would also like to thank K L Bhanu Moorthy for the discussions.

%\section{Discussion}

%\section{Future work}

%\section{Conclusions}

%\section{Acknowledgements}

\bibliographystyle{IEEEtran}

\bibliography{mybib}

\end{document}